\documentclass[twoside,journey]{IEEEtran}
\usepackage{makecell}
\usepackage{hyperref}
\usepackage{array}
\usepackage{graphicx,amssymb,amsmath}
\usepackage{multicol}
\usepackage[noadjust]{cite}
\usepackage{setspace}
\usepackage{subfigure}
\usepackage{float}
\usepackage{url}
\usepackage{stfloats}
\usepackage{amsthm,pifont}
\usepackage{flushend}
\usepackage{cases,subeqnarray}
\usepackage{bm,multirow,bigstrut}
\usepackage{amsmath, amsthm, amssymb}
\usepackage{textcomp}
\usepackage{latexsym,bm}
\usepackage{booktabs}
\usepackage{xcolor}
\usepackage{mathtools}
\usepackage{dsfont}
\usepackage{extarrows}
\usepackage{epsfig}
\usepackage{epstopdf}
\usepackage{colortbl}
\usepackage[noend]{algpseudocode}
\usepackage{algorithmicx,algorithm}
\theoremstyle{plain}

\theoremstyle{plain}

\usepackage{amsmath}

\definecolor{Gray}{gray}{0.85}
\IEEEoverridecommandlockouts
\begin{document}

%----------------------------title&author&thanks----------------------------
% \title{When Networks Meets Generative AI: Potentials, Challenges, and Directions}
\title{The Age of Generative AI and AI-Generated Everything}
\author{Hongyang~Du, Dusit~Niyato,~\IEEEmembership{Fellow,~IEEE}, Jiawen~Kang, Zehui~Xiong, Ping~Zhang,~\emph{Fellow,~IEEE}, Shuguang~Cui,~\emph{Fellow,~IEEE}, Xuemin~(Sherman)~Shen,~\emph{Fellow,~IEEE}, Shiwen~Mao,~\emph{Fellow,~IEEE}, Zhu~Han,~\emph{Fellow,~IEEE}, Abbas~Jamalipour,~\emph{Fellow,~IEEE}, H.~Vincent~Poor,~\emph{Life~Fellow,~IEEE}, and Dong~In~Kim,~\emph{Fellow,~IEEE}
\thanks{H.~Du and D.~Niyato are with the School of Computer Science and Engineering, Nanyang Technological University, Singapore 639798, Singapore (e-mail: hongyang001@e.ntu.edu.sg; dniyato@ntu.edu.sg). J.~Kang is with the School of Automation, Guangdong University of Technology, and Key Laboratory of Intelligent Information Processing and System Integration of IoT, Ministry of Education, Guangzhou 510006, China, and also with Guangdong-HongKong-Macao Joint Laboratory for Smart Discrete Manufacturing, Guangzhou 510006, China (e-mail: kavinkang@gdut.edu.cn). Z.~Xiong is with the Pillar of Information Systems Technology and Design, Singapore University of Technology and Design, Singapore 487372, Singapore (e-mail: zehui\_xiong@sutd.edu.sg). P. Zhang is with the State Key Laboratory of Networking and Switching Technology, Beijing University of Posts and Telecommunications, China (e-mail: pzhang@bupt.edu.cn). S. Cui is with the School of Science and Engineering (SSE) and the Future Network of Intelligence Institute (FNii), Chinese University of Hong Kong (Shenzhen), China (e-mail: shuguangcui@cuhk.edu.cn). X.~Shen is with the Department of Electrical and Computer Engineering, University of Waterloo, Waterloo, ON N2L 3G1, Canada (e-mail: sshen@uwaterloo.ca). S.~Mao is with the Department of Electrical and Computer Engineering, Auburn University, Auburn, AL 36849-5201 USA (email: smao@ieee.org). Z.~Han is with the Department of Electrical and Computer Engineering, University of Houston, Houston, TX 77004 USA, and also with the Department of Computer Science and Engineering, Kyung Hee University, Seoul 446-701, South Korea (e-mail: hanzhu22@gmail.com). A.~Jamalipour is with the School of Electrical and Information Engineering, University of Sydney, Sydney, NSW 2006, Australia (e-mail: a.jamalipour@ieee.org). H.~V.~Poor is with the Department of Electrical and Computer Engineering, Princeton University, Princeton, NJ 08544, USA (e-mail: poor@princeton.edu). D.~I.~Kim is with the Department of Electrical and Computer Engineering, Sungkyunkwan University, Suwon 16419, South Korea (email:dikim@skku.ac.kr).}
}
\maketitle
%----------------------------abstract----------------------------
\vspace{-1cm}

\begin{abstract}
Generative AI (GAI) has emerged as a significant advancement in artificial intelligence, renowned for its language and image generation capabilities. 
This paper presents ``AI-Generated Everything'' (AIGX), a concept that extends GAI beyond mere content creation to real-time adaptation and control across diverse technological domains.
In networking, AIGX collaborates closely with physical, data link, network, and application layers to enhance real-time network management that responds to various system and service settings as well as application and user requirements. Networks, in return, serve as crucial components in further AIGX capability optimization through the AIGX lifecycle, i.e., data collection, distributed pre-training, and rapid decision-making, thereby establishing a mutually enhancing interplay.
Moreover, we offer an in-depth case study focused on power allocation to illustrate the interdependence between AIGX and networking systems.
Through this exploration, the article analyzes the significant role of GAI for networking, clarifies the ways networks augment AIGX functionalities, and underscores the virtuous interactive cycle they form. This article paves the way for subsequent future research aimed at fully unlocking the potential of GAI and networks. 
\end{abstract}
%----------------------------keywords----------------------------
\newpage
\IEEEpeerreviewmaketitle
%----------------------------introduction----------------------------
\section{Introduction}
\IEEEPARstart{W}{ithin} the evolving field of artificial intelligence (AI), the shift is evident from merely analyzing expansive datasets to actively generating innovative content. Milestones like AlphaGo's 2016 victory over a Go world champion laid the foundation, but Generative AI (GAI) represents a significant turn in AI's progression~\cite{du2023beyond}.
ChatGPT exemplifies this trend with its advanced conversational capabilities, resulting in more context-aware and detailed user interactions~\cite{baidoo2023education}. 
Similarly, DALL-E 3's ability to produce images from text descriptions blends linguistic comprehension with visual creation.
Such GAI advancements underscore the expanding role of machines in domains once believed to be unique to human creativity. 
This growth in GAI promises to reshape our view of AI across academic, industrial, and societal spheres~\cite{xu2023unleashing}.
\begin{figure*}[t!]
\centering
\includegraphics[width=0.9\textwidth]{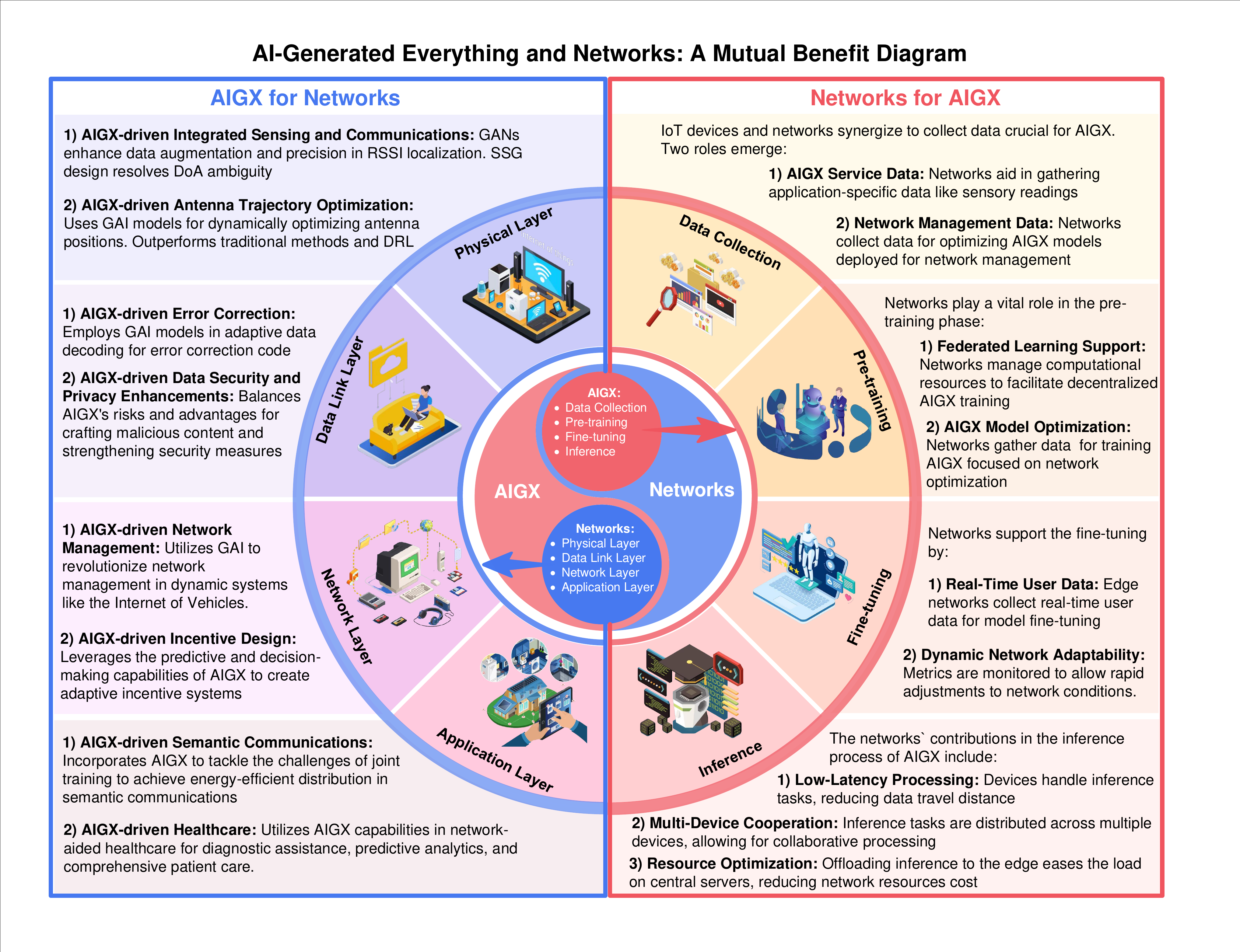}
\caption{Symbiotic interaction between AIGX and data communications and networking. On the left, the diagram delineates the functionality of AIGX across various network layers: physical, data link, network, and application layers. On the right, it illustrates how networks contribute at distinct stages of AIGX's lifecycle, including data collection, pre-training, fine-tuning, and inference.}
\label{images111}
\end{figure*}

AI-driven networks have conventionally employed Discriminative AI (DAI) models, adept at tasks like data classification and prediction~\cite{zuo2023survey}. While DAI excels at detecting existing patterns, GAI extends capabilities by creating new data samples, for instance, entirely fresh images or audio not present in original datasets. This introduces broadened applications in content creation, data augmentation, and even crafting network optimization strategies~\cite{du2023beyond}. This capability transition positions GAI as a pivotal tool in network functions:
\begin{itemize} 
\item \textbf{Data Synthesis and Augmentation:} Beyond DAI's data interpretation, GAI creates synthetic data vital for humans and networks. This is particularly useful in simulating network traffic and generating test data. Especially with limited datasets, GAI synthesizes samples of rare network faults and attacks to enhance anomaly-detection training.
\item \textbf{Predictive Analysis and Management:} More than just identifying patterns as DAI does, GAI generates predictive actions according to network condition changes using past data and forthcoming events. This proactive approach ensures efficient resource distribution and minimizes potential network congestion.
\item \textbf{Personalized User Interaction:} GAI refines network-driven services by personalizing user experiences, enriching content, formulating recommendations, and designing network interfaces that align with individual user preferences, surpassing what DAI offers.
\end{itemize}
As we embrace these merits, merging GAI into networking and technical areas presents unique challenges and opportunities. Within this confluence of GAI and networks emerges the innovative notion of AI-Generated Everything (AIGX).

\textit{AIGX, progressing beyond AIGC's human-oriented content creation, represents a paradigm that utilizes GAI to devise, refine, and optimize applications and systems, enabling them to interact and adapt to instantaneous environmental shifts dynamically.} Spanning from reshaping transportation and healthcare to innovating in urban planning and optimizing power grids, AIGX promises widespread transformation. Particularly in networking, a field ripe for AIGX-driven evolution, GAI's influence on every network component—from content delivery to architectural configurations—is pivotal~\cite{xu2023unleashing}. AIGX enables dynamic adaptations to real-time conditions~\cite{du2023beyond}, deploys predictive insights for improved decision-making~\cite{du2023generative}, and introduces resource allocation schemes that ensure optimal performance~\cite{du2023ai}. 

While AIGX aims to revolutionize networks, the networks reciprocally play a significant role in optimizing AIGX. This symbiosis is evident throughout the AIGX lifecycle, including data collection, pre-training, fine-tuning, and inference. Specifically, the Internet-of-Things (IoT) is pivotal for efficiently collecting and preprocessing massive data streams from interconnected devices~\cite{xiao2022edge}. In the training phase, federated learning (FL) becomes useful, providing a decentralized AIGX model training paradigm that avoids centralized data hub limitations~\cite{huang2023federated}. This approach facilitates localized data processing at the edge and enhances training and inference. Meanwhile, specialized offloading strategies ensure computational tasks are intelligently distributed from resource-constrained devices to more resourceful nodes or cloud infrastructures~\cite{du2023exploring}.
\begin{figure*}[t!]
	\centering
	\includegraphics[width=1\textwidth]{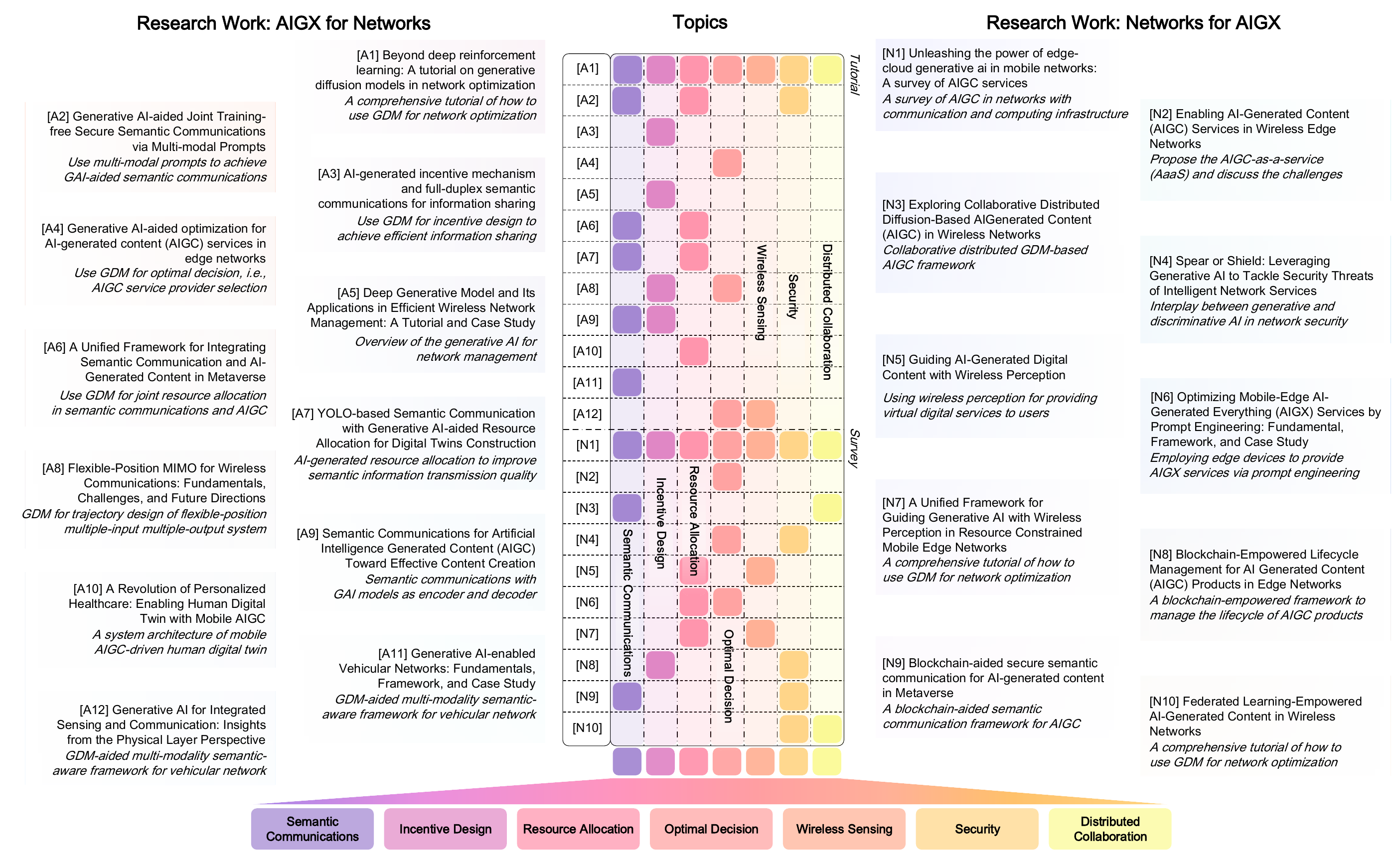}
	\caption{Summary of research topics and key contributions in {\textit{AIGC for networks}} and {\textit{networks for AIGX}}.}
	\label{faef}
\end{figure*}
As shown in Fig.~\ref{faef}, recent advances in AIGX highlight its symbiotic evolution with intelligent networks. This paper elucidates this collaboration and its consequences for computing and communication frameworks. Our main contributions are:
\begin{itemize}
	\item \textit{AIGX Enhancing Networks:} We discuss the impacts of AIGX across network layers and show how AIGX facilitates real-time modulation adjustments, elevates data security, and tailors adaptive resource management in various network settings.
	
	\item \textit{Networks Enabling AIGX:} We spotlight the role of networks throughout the AIGX lifecycle. From enabling efficient data collection and distributed learning during pre-training stages to contributing to AIGX model optimization and ensuring low latency during the inference process, networks emerge as pivotal to deploying AIGX capabilities.
	
	\item \textit{virtuous interactive cycle:} Through a case study focused on power allocation, we examine the ``virtuous interactive cycle'' between AIGX and networks. This exploration not only exemplifies the symbiotic relationship between AIGX and intelligent networks in real-world scenarios but also serves to offer tangible guidelines for the integration of AIGX techniques into network design.
\end{itemize}

\section{Generative AI and AI-Generated Everything}
In this section, we delve into the foundational techniques of GAI and also introduce the AIGX paradigm.

\subsection{Generative AI: Core Techniques}
GAI and DAI hold distinct methodologies and proficiencies. While DAI predominantly focuses on distinguishing between inputs by outlining class boundaries, GAI emphasizes generating content reminiscent of its training data.

\subsubsection{Generative Adversarial Networks (GANs)}
GANs are pivotal models within GAI, expertly integrating data generation and discrimination capabilities. The model consists of two key components: a generator for data production and a discriminator to distinguish between original and generated data. Engaging in iterative training and competition, the generator crafts increasingly accurate data. In networking, GANs enable the simulation of complex network traffic patterns for cybersecurity, enriching datasets to train sophisticated intrusion detection systems.

\subsubsection{Transformers}
Initially developed for natural language processing, Transformers have broadened their impact, notably influencing the GAI. Their core advantage lies in attention mechanisms, which assign varied importance to different input sections, enabling effective parallel processing. In networking, Transformers provide real-time bottleneck prediction and optimize packet routing based on historical traffic trends. Models like BERT and ChatGPT, which excel in linguistic applications, find use in designing encoders and decoders for semantic communications (SemCom)~\cite{xu2023unleashing}.

\subsubsection{Generative Diffusion Models (GDMs)}
GDMs represent a novel facet of GAI, gradually converting an initial random sample into a targeted output through multiple iterative denoising steps. These models offer a departure from traditional neural architectures, presenting a new approach to content generation. In networking, GDMs have been applied in various network optimization tasks such as resource allocation, error correction coding, network economics, and SemCom~\cite{du2023beyond}.

\subsubsection{Other GAI Techniques}
{\textit{Autoregressive Models (ARMs)}} leverage sequences of previous values to predict upcoming data points and are notably effective in generating sequential content such as text, audio, or video. ARMs can forecast network loads, optimize bandwidth allocation, and identify potential failure points using historical data in networks. {\textit{Variational Autoencoders (VAEs)}} capture compressed data representations, generating new samples from this latent space, and are vital in creating synthetic network traffic patterns or enhancing multimedia content to boost user Quality of Experience (QoE) in content delivery frameworks. {\textit{Flow-based Models (FBMs)}}, on the other hand, facilitate data generation by converting basic distributions into complex target ones through reversible transformations, ensuring content aligns with specified distributions, which is crucial for dynamic content stream adjustments and network security tasks where accurate traffic pattern replication is vital.
%\subsubsection{Autoregressive Models (ARMs)}
%ARMs forecast future data points using a sequence of prior values. In the context of GAI, ARMs excel in generating sequential content, encompassing text, audio, or video. ARMs can be used within network settings to project network loads, allocate bandwidth, or anticipate potential failure sites informed by historical network performance data.
%
%\subsubsection{Variational Autoencoders (VAEs)}
%VAEs derive a compressed representation of input data, enabling the generation of novel data samples from this latent space. With two primary components, the encoder, and the decoder, VAEs facilitate the transition from input to synthesized content. VAEs can be utilized for generating synthetic network traffic patterns to train models. Additionally, they may produce supplementary multimedia content to enhance user quality of experience (QoE) within content delivery networks.
%
%\subsubsection{Flow-based Models (FBMs)}
%FBMs enhance data generation by morphing elemental distributions into complex target distributions via reversible transformations. Leveraging neural architectures for these transformations ensures that the generated content aligns with specified distributions. In networking, FBMs are valuable for dynamically adapting content streams, and in network security tasks where accurate replication of specific traffic patterns is crucial.

\subsection{AI-Generated Everything (AIGX): The New Network Paradigm}
\subsubsection{Defining AIGX, An Evolution from AIGC}
AIGC refers to the application of AI techniques to facilitate and automate the creation of content that is specifically tailored to user preferences and requirements. The `C' in AIGC emphasizes the content-centric nature of this application, wherein the generated outputs are primarily forms of media or information such as text, images, videos, 3D models, and audio~\cite{du2023exploring}.

AIGX builds upon the foundational concepts of AIGC but extends its reach. `X' in AIGX denotes `Everything', representing its influence across all technological aspects. Instead of just generating content, AIGX leverages GAI to adaptively design, fine-tune, and optimize applications and systems as they interact with real-time environmental changes. For instance, in telecommunications, AIGX can dynamically adjust network resources for maximal user QoE. In manufacturing, AIGX could redefine automation by enabling machines to detect and proactively manage production anomalies.

\subsubsection{Virtuous Interactive Cycle of AIGX and Networks}
Embedding AIGX into network systems symbolizes transitioning from traditional static architectures to a dynamic, GAI-driven framework. Specifically, AIGX and networks have the following reciprocal benefits:
\begin{itemize}
	\item {\textbf{AIGX for Networks.}} AIGX facilitates enhancements across network layers. In the {\textit{Physical Layer}}, it enables real-time modulation adjustments, optimizing data rates and minimizing errors based on channel conditions~\cite{du2023beyond}. In the {\textit{Data Link Layer}}, AIGX dynamically augments error correction algorithms adapting to interference levels and enhances data security~\cite{choukroun2022denoising,xu2023unleashing}. In the {\textit{Network Layer}}, it improves management in specialized systems, such as the Internet of Vehicles (IoV), and generates adaptive incentive mechanisms~\cite{du2023ai}. At the {\textit{Application Layer}}, AIGX benefits the design of SemCom and healthcare systems~\cite{du2023generativemul}.
	\item {\textbf{Networks for AIGX.}} Network infrastructure is pivotal across all AIGX lifecycle stages. During data collection, networks enable the collection of both application-specific and network management data, fostering a feedback loop that boosts AIGX capabilities and network efficiency. In the pre-training phase, FL, supported by the network, facilitates decentralized model training, further optimizing AIGX models~\cite{huang2023federated}. The fine-tuning stage leverages networks for real-time data collection and dynamic adaptability~\cite{huang2023federated}. Lastly, networks employ edge offloading and multi-device cooperation during inference, reducing latency and optimizing resource allocation to enhance system throughput~\cite{du2023exploring}.
\end{itemize}
This interplay between AIGX and networks marks a paradigm shift, transforming networks into dynamic, evolving ecosystems that continuously adapt to immediate needs. Next, we delve into two aspects of the mutually beneficial relationship between AIGX and networks, followed by a case study to illustrate this virtuous interactive cycle and gains.

\section{AIGX for Networks}
This section studies AIGX methodologies that permeate and influence the network architecture's layers.
\begin{figure*}[t!]
	\centering
	\includegraphics[width=0.9\textwidth]{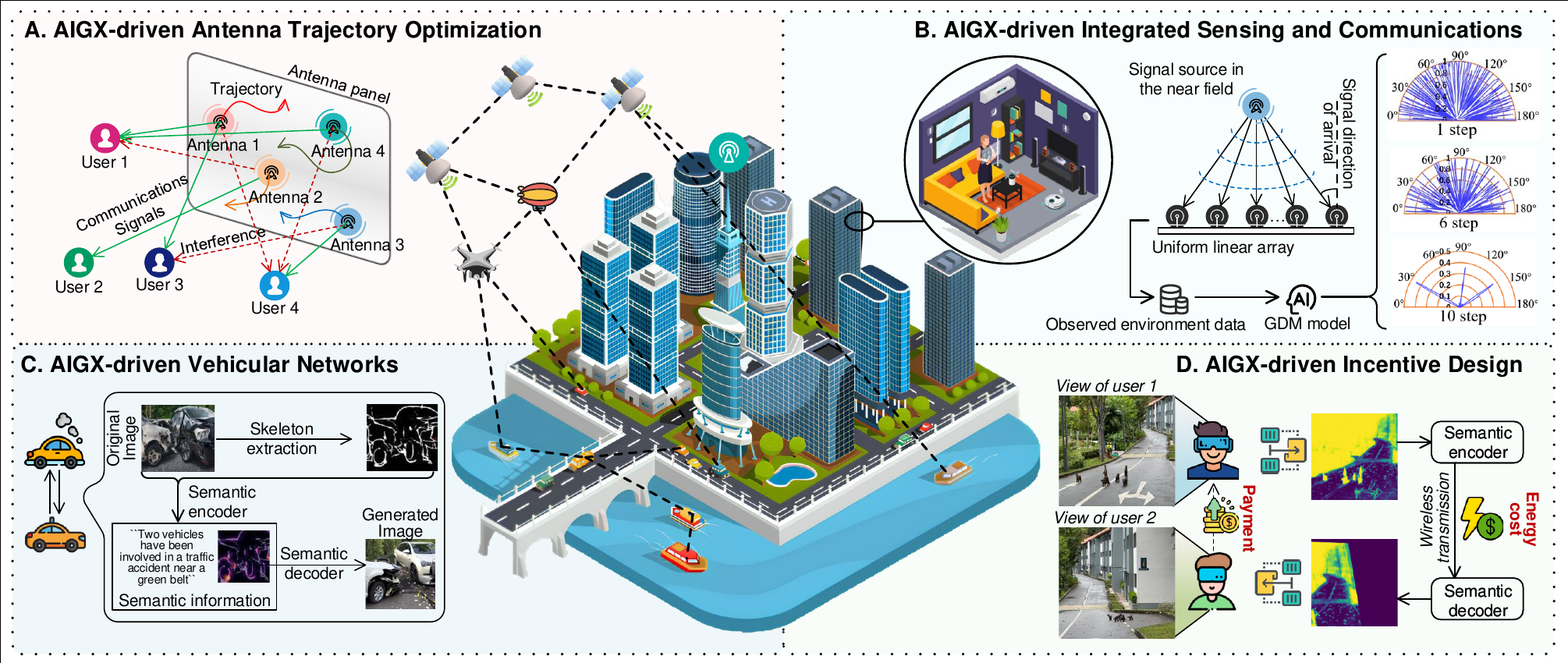}
	\caption{Representative system models of AIGX technologies in networks. {\textit{Part A}} depicts the generation of optimal antenna trajectories using AIGX. {\textit{Part B}} demonstrates the localization results produced by AIGX, leveraging perceived wireless environments. {\textit{Part C}}, the application of AIGX technology for designing encoders and decoders in vehicular network semantic communication systems is presented. {\textit{Part D}} shows the incentive mechanism generated by AIGX in mixed-reality user information-sharing systems.}
	\label{imagesPhysicalLayer}
\end{figure*}

\subsection{Physical Layer}

\subsubsection{AIGX-driven Integrated Sensing and Communications (ISAC)}
ISAC merges wireless sensing and communication to efficiently use constrained resources, with applications ranging from autonomous driving to gesture identification~\cite{liu2022survey}. Integrating GAI with ISAC leads to novel applications and enhancements. 
In ISAC data processing, GAI-powered tools like GANs amplify system accuracy using limited real-world data, particularly evident in Received Signal Strength Indicator (RSSI) fingerprint localization. GAI adoption enhances dataset diversity, bolstering human activity detection precision. More complex issues, such as signal Direction of Arrival (DoA) estimation in both near- and far-field ISAC systems, benefit from GAI's capability to solve problems like phase ambiguity. For instance, a GDM-developed signal spectrum generator (SSG) is proposed in {\textbf{{\textit{[A12]}}}} of Fig.~\ref{faef}, as shown in Part B of Fig.~\ref{imagesPhysicalLayer}. In a test with four antennas, the SSG used 10,000 paired signal spectrums, designated $80\%$ for training and $20\%$ for validation, and introduced noise into expert-generated solutions, followed by a stepwise denoising process. The evaluation revealed that the SSG output converges with expert solutions during training and achieves a test loss of $-10$, outperforming the $-80$ loss in deep reinforcement learning (DRL)-based methods. 
% This underscores the SSG's enhanced focus on spectral elements corresponding to DoAs and enables robust learning of accurate solutions.

\subsubsection{AIGX-driven Antenna Trajectory Optimization}
With their repositionable antennas, flexible-position MIMO systems improve wireless communications by optimizing channel conditions and boosting spectral and energy efficiencies. The key challenge is trajectory optimization for maximizing spectral (SE) or energy efficiency (EE).
Traditional optimization techniques can get stuck in complex scenarios, often settling in local optima. In contrast, AIGX employs GAI-driven optimization using GDMs for dynamic network adjustments~\cite{du2023beyond}.
In the case study in {\textbf{{\textit{[A8]}}}} of Fig.~\ref{faef}, a GDM aimed to enhance SE is trained to generate antenna trajectories, as shown in Part A of Fig.~\ref{imagesPhysicalLayer}. 
Compared to DRL methods, which quickly peak and then level out, AIGX optimization steadily increases rewards, raising the sum SE from an average of $11.7$ (using DRL) to $13.3$. Results revealed that the generated solution enabled antennas to either adjust positions to enhance user coverage or move to the system's periphery to mitigate multi-user interference to optimize SE, while EE-prioritized scenarios prompted antennas to shift towards areas with higher user density.

\subsection{Data Link Layer}
\subsubsection{AIGX-driven Error Correction}
Error correction is significant for ensuring data integrity across interference-prone channels in the data link layer. Identifying the codeword most suitable with the received signal is traditionally considered an NP-hard challenge, implying a potential exponential search for optimal decoding.
Fortunately, GAI models in the AIGX framework, like Transformer-based decoders, have brought efficiency to this task, enhancing accuracy and computational speed~\cite{choukroun2022denoising}. 
However, the invariant computational load is a persistent challenge, independent of the codeword corruption degree. 
By introducing the GDM as a solution, AIGX approaches decoding iteratively according to the varied codeword corruption to reduce computational demands significantly~\cite{choukroun2022denoising}. 
A case study in~\cite{choukroun2022denoising} highlights AIGX's effectiveness. It presents data transmission as an iterative forward diffusion process that requires inversion at the receiving end. 
The Bit Error Rate (BER) of the GDM method is notably lower than traditional schemes, with it being only $11\%$ of a specific Transformer scheme when the signal-to-noise ratio is at $4$ dB.

\subsubsection{AIGX-driven Data Security and Privacy Enhancements}
The surge in digital communication has heightened the importance of effective security and privacy measures. While AIGX introduces novel ways to create content, it also opens doors to potential risks. We discuss the attack scenarios and defense mechanisms as:
\begin{itemize}
	\item \textbf{Attack Scenarios}: Powerful models, like Large Language Models (LLMs), including OpenAI's ChatGPT, introduce a new type of threat. Such models can generate harmful content that bypasses safeguards put in place by API providers, as discussed in {\textbf{{\textit{[N4]}}}} of Fig.~\ref{faef}. Additionally, AIGX might exploit subtle similarities between encrypted and plain images in the embedding space, compromising the trustworthiness of prevailing encryption methods.
	\item \textbf{Defense Mechanisms}: LLMs can enhance the training dataset of a DAI model by creating adversarial examples. Take a network intrusion detection system (NIDS) trained to distinguish between `normal' and `malicious' network traffic. If a malicious pattern, like several failed logins followed by a successful one, is detectable, the LLM might generate an adversarial sample that spreads the login attempts over various user accounts or IPs. This disrupts the recognizable pattern and might go unnoticed.
\end{itemize}
An illustrative example of the synergy between AIGX's advantages and associated risks is evident in wireless image transmission. In this setting, discriminative AI attempts to disrupt communication by initiating data poisoning attacks on an image dataset hosted on a server. Counteracting this, AIGX-driven defenses utilize a GDM to authenticate each image before transmission. As shown in {\textbf{{\textit{[N4]}}}} of Fig.~\ref{faef}, utilizing this AIGX-driven defense led to an $8.7\%$ reduction in energy consumption. 
% Moreover, the need for retransmission due to erroneous content drastically declined, from an initial $32$ images without any defense to a significantly lower count of just $6$.

\subsection{Network Layer}
\subsubsection{AIGX-driven Network Management}
Network management is crucial for modern systems, ensuring efficient communication and performance, especially in extensive networks with data flow across many devices~\cite{song2023aerial}. AIGX, with GAI's capability to discern and emulate intricate data patterns, enhances troubleshooting, predictive maintenance, and data synthesis~\cite{du2023beyond}.
Consider the IoV, where vehicles continuously exchange data. The dynamism of this system requires adaptable network management. AIGX, given its data representation and generation prowess, can predict network congestion, allocate bandwidth adaptively, and even fill in data gaps where actual data is lacking. As shown in Part C of Fig.~\ref{imagesPhysicalLayer}, a specialized study, i.e., {\textbf{{\textit{[A11]}}}} of Fig.~\ref{faef}, addresses V2V resource allocation, formulating a QoE metric grounded in transmission rate and received image fidelity. Comparative analyses reveal that AIGX-based approaches outperform DRL counterparts, recording an $18.5\%$ increase in average QoE.
 % when image payload sizes are $7$ units

\subsubsection{AIGX-driven Incentive Design}
Incentives play a key role in prompting network participants to share resources, boosting network efficiency. Without the right rewards, users might hold back, considering costs like battery usage or bandwidth.
Unlike the conventional DAI-based method, AIGX models adjust to changing network conditions and user behaviors, offering a way to design smarter, adaptive incentives.
As shown in Par D of Fig.~\ref{imagesPhysicalLayer}, a noteworthy application involves deploying AIGC within Mixed-Reality (MR) technologies~\cite{du2023ai}. MR headset-mounted devices (HMDs) often face constraints in computational power, impacting user experience. 
An effective information-sharing strategy, leveraging full-duplex device-to-device (D2D) SemCom, emerges as a solution~\cite{du2023ai}. 
Instead of each user doing repetitive computational tasks like generating AIGC, the system facilitates sharing this content with relevant semantic information to nearby users.
Such a framework calls for an effective incentive mechanism to motivate users. An AIGX-based incentive model rooted in contract theory has been proven to create optimal agreements benefiting all. It boosts user experience quality by $11.7\%$ over the traditional DRL approach~\cite{du2023ai}.

\subsection{Application Layer}
\subsubsection{AIGX-driven Semantic Communications}
Semantic Communications (SemCom) addresses the challenge of exponential data volume growth in wireless communication networks by converting messages into semantic information for transmission using a semantic encoder and decoder. However, challenges exist in joint training and energy-efficient distribution of these AI-based encoders and decoders. Fortunately, GAI models, such as advanced language and image generation models, can reconstruct complex messages from simpler semantic representations, alleviating the need for joint training~\cite{han2023generative}. 
For example, using multi-modal prompts, i.e., visual and textual prompts, can lead to accurate semantic decodings~\cite{du2023generativemul} to solve the problem of instability brought by the diverse generative capabilities of GAI models, which can be used in scenarios that require accurate information transfer such as human face images. 
Another example is integrating SemCom and AIGC (ISGC) to enhance user immersion as discussed in {\textbf{{\textit{[A6]}}}} of Fig.~\ref{faef}. ISGC balances computing and communication resources for semantic extraction, AIGC inference, and graphic rendering. An effective resource allocation mechanism can be achieved with the help of AIGX to obtain near-optimal strategies. Numerical results show that the GDM-based method can improve the QoE by $8.3\%$ compared with the Proximal Policy Optimization (PPO) method.

\subsubsection{AIGX-driven Healthcare}
Network technologies, aided by IoT, are transforming healthcare, allowing real-time patient monitoring and efficient data sharing. AIGX further elevates network-based healthcare by simulating human thought and analyzing large datasets, improving diagnostics, predictions, and overall care, as discussed in {\textbf{{\textit{[A10]}}}} of Fig.~\ref{faef}. It can foresee health concerns by reviewing vast data and devising personalized treatment plans using a patient's history and present health data. In AIGX-driven health applications like virtual physical (VR) therapy, maintaining high-quality VR video streams is essential. Although SemCom helps reduce data, preserving VR quality is a challenge, especially with potential inaccuracies. AIGX can recreate these streams closely to the original. Yet, managing the network's efficiency, VR video quality, and genuineness can be challenging.
An optimization design is presented in {\textbf{{\textit{[A10]}}}} of Fig.~\ref{faef}, considering constraints like bandwidth, computational capacity, and QoE benchmarks, aiming to enhance user QoE by focusing on aspects including resolution ratio and the diffusion step that is crucial for GDMs. Compared to the Soft Actor-Critic (SAC) algorithm, a conditional GDM-based approach has increased the total users' QoE by $32.4\%$.

\subsection{Lesson Learned}
In leveraging AIGX methodologies for network systems, several key lessons emerge. The primary motivation behind leveraging GAI is its ability to amplify accuracy, optimize network management schemes, and predict intricate data patterns, especially in dynamic systems like the IoV.
Conventional DRL-based methods, prone to getting stuck in local optima, can struggle with real-time changes. GAI, using tools such as GANs and GDMs, provides flexible solutions that adapt to evolving conditions. 
This flexibility is crucial when navigating vast, dynamic systems or addressing challenges like error correction. 
Furthermore, while AIGX's introduction brings forth data security opportunities, it demands careful implementation. 
In practical applications, areas like healthcare are transformed by AIGX-driven networks, facilitating real-time tracking, individualized treatments, and enhanced diagnostic capabilities.

\section{Network for AIGX}
Networks play a pivotal role in the AIGX lifecycle, from data collection and training to fine-tuning and inference. This section delves into each of these aspects.
\begin{figure*}[t!]
	\centering
	\includegraphics[width=1\textwidth]{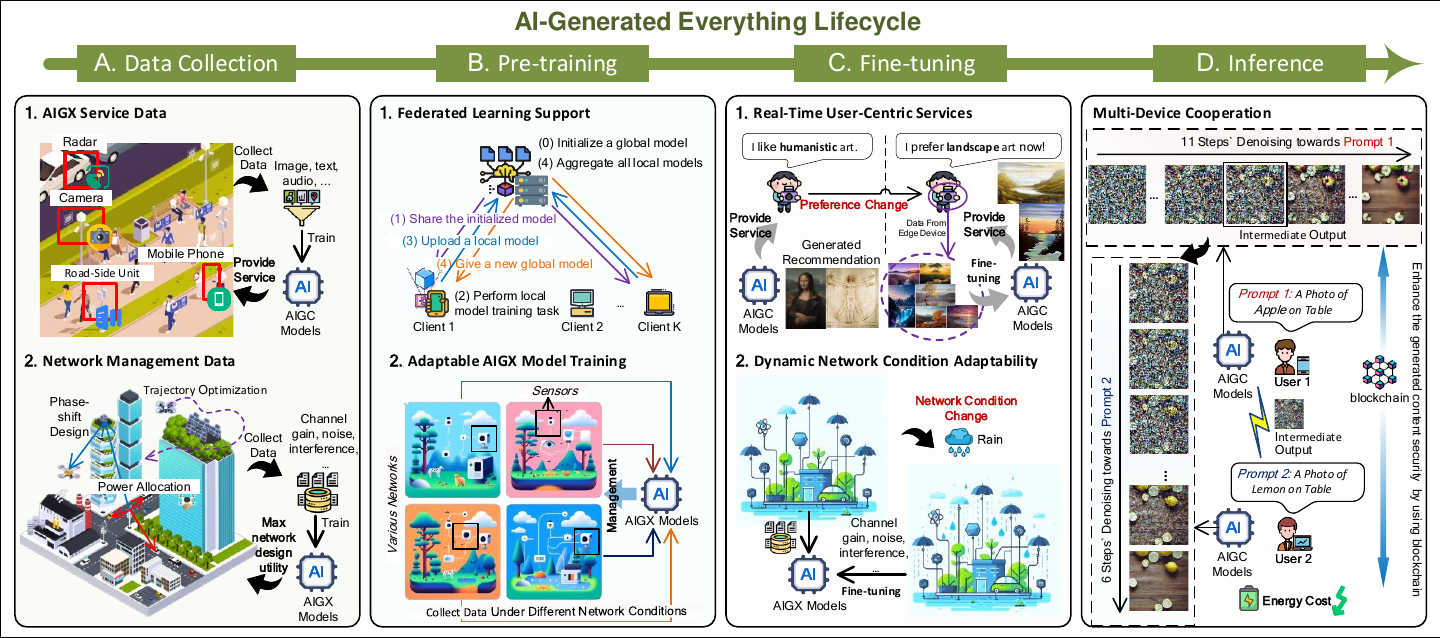}
	\caption{The role of networks across various stages of the AIGX lifecycle. In the {\textit{Data Collection}} phase, networks enable efficient data gathering essential for training effective AIGX models that cater to user needs or manage the network. During the {\textit{Pre-training}} stage, networks support AIGX models in flexible training, such as FL supporting distributed training. In the {\textit{Fine-tuning}} stage, network devices update pre-trained models with new user data for personalized services and adapt decision-making models to emerging network conditions. Collaboration among network devices at the {\textit{Inference}} stage promotes more adaptable and energy-efficient inference models.}
	\label{netforaigc}
\end{figure*}
\subsection{Data Collection}
Data collection is fundamental to the AIGX ecosystem, impacting the efficiency and reliability of AIGX applications. Utilizing varied techniques enriched by IoT capabilities, integrated devices, and sensors are crucial channels for detailed environmental data collection. In smart cities, for instance, IoT devices acquire data on air quality, vehicular movement, and energy consumption, and use these data for AIGX applications. Networks play two synergistic roles in AIGX data collection:
\begin{itemize}
	\item \textbf{AIGX Service Data:} As shown in Part A.1 of Fig.~\ref{netforaigc}, networks enable the acquisition of application-specific data, such as sensory and visual information from IoT cameras, which form the empirical foundation for developing and refining AIGX service models.
	\item \textbf{Network Management Data:} Beyond service-specific needs, networks collect data useful for fine-tuning and optimizing AIGX models deployed for network management. As shown in Part A.2 of Fig.~\ref{netforaigc}, expert decisions under various channel conditions can be collected for further AIGX model training.
\end{itemize}
This dynamic fosters a virtuous interactive cycle: AIGX models optimized for network performance enhance data rates and network efficiency, which in turn improve data collection for AIGX model refinement, perpetually fine-tuning both AIGX capabilities and network efficacy.

\subsection{Pre-training}
The pre-training phase in the AIGX lifecycle, crucial for developing foundational models across various applications, significantly leverages network infrastructure capabilities: 
\begin{itemize}
	\item \textbf{Federated Learning Support:} Networks employ FL, allowing decentralized model training while keeping data at the edge~\cite{huang2023federated} as shown in Part B.1 of Fig.~\ref{netforaigc}. While FL enhances data privacy, it necessitates effective network resource management to address computational challenges. Networks manage resources, including dynamic bandwidth allocation and low-latency communication, ensuring efficient localized AIGX model training across devices without overloading the network~\cite{huang2023federated}. 
	\item \textbf{Adaptable AIGX Model Training:} Networks, by collecting diverse environmental and network-related data such as latency and bandwidth utilization, enrich AIGX model training input as shown in Part B.2 of Fig.~\ref{netforaigc}, focusing on network optimization. For instance, if a pre-trained AIGX model incorporates high-bandwidth environment data, its performance may be suboptimal under bandwidth constraints.
\end{itemize}
This symbiotic relationship ultimately achieves improvements in both AIGX functionalities and network performance.

\subsection{Fine-tuning}
In contrast to the generalizations of the pre-training stage, fine-tuning zeroes in on swift optimization for user preferences and fluctuating network conditions. The network bolsters AIGX fine-tuning as follows:
\begin{itemize}
	\item \textbf{Real-Time User-Centric Services:} Leveraging the capabilities of edge networks, as shown in Part C.1 of Fig.~\ref{netforaigc}, real-time data reflecting user preferences is collected for model fine-tuning~\cite{huang2023federated}. This makes AIGX models agile in adapting to shifts in user behavior or network dynamics.
	\item \textbf{Dynamic Network Condition Adaptability:} For network management-focused AIGX models, the network continuously monitors metrics like fadings and packet loss, feeding this data into the fine-tuning process to swiftly adapt to network changes as shown in Part C.2 of Fig.~\ref{netforaigc}.
\end{itemize}
Consequently, the network functions as an effective platform, fostering the rapid fine-tuning of AIGX models, and making them practical for real-world scenarios.
%It enables the collection of real-time data to meet shifts in user preferences for service-specific applications and dynamically changing network conditions for management tasks. 

\subsection{Inference}
The inference stage deploys trained AIGX models for specific needs. Rather than depending on traditional centralized servers, which often result in bottlenecks and delays, AIGX can leverage edge-based offloading to enhance the efficiency~\cite{du2023exploring}. The network's role in AIGX inference includes:
\begin{itemize}
	\item \textbf{Low-Latency Processing:} Offloading inference tasks to edge devices significantly minimizes latency. This is achieved by reducing the round-trip data travel distance between the user and the processing unit, thereby enhancing real-time responsiveness.
	\item \textbf{Multi-Device Cooperation:} AIGX inference can be partitioned and executed across multiple cooperating edge devices as shown in Part D of Fig.~\ref{netforaigc}. This distributed method removes individual device limits, including computation and energy resources, and boosts the overall system effectiveness through collaborative processing.
	\item \textbf{Optimized Resource Allocation:} Distributing the inference process across edge devices alleviates central server workloads, optimizing the utilization of network resources and averting potential congestion points.
\end{itemize}
With edge offloading and multi-device cooperation, the network sharply reduces latency, enhances system throughput, and optimizes resource allocation.

\subsection{Lesson Learned}
Networks are integral to the AIGX ecosystem, providing the infrastructure and data pathways that support AIGX models. By enabling decentralized training, networks ensure data privacy while enhancing model adaptability, essential in ever-changing real-world scenarios. 
Networks are pivotal in gathering diverse data, which strengthens AIGX models, and are indispensable for ensuring real-time adaptability to user preferences and shifting network conditions. 
Furthermore, with the trend towards edge-based offloading, networks optimize resource allocation and reduce latency, ensuring that AIGX models are both efficient and practical in real-world applications.

\section{Case Study: The virtuous interactive cycle of AIGX and Networks in Power Allocation}
\begin{figure*}[t!]
	\centering
	\includegraphics[width=0.95\textwidth]{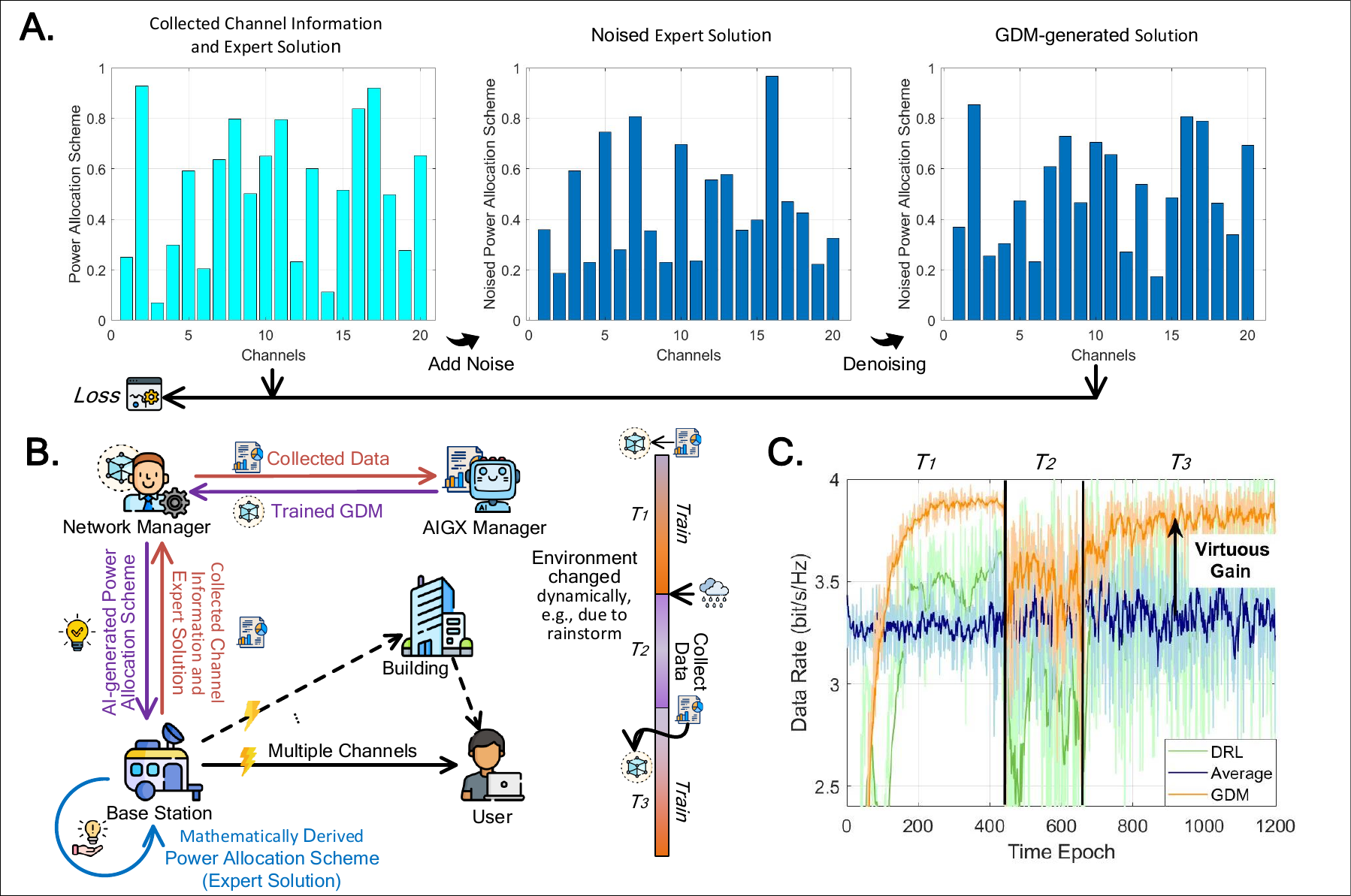}
	\caption{Case study of the AIGX-network virtuous interactive cycle: {\textit{Part B}} shows the system model where a base station and user communicate through multiple channels, demanding power allocation to optimize the user's sum rate. Networks feed training data to the AIGX Manager, which generates a power allocation model. {\textit{Part A}} shows the training process of the GDM, generating optimal decisions under given channel conditions by adding noise to the expert solution and denoising it. {\textit{Part C}} reveals the AIGX-network cycle's adaptation to network changes, illustrating how improved sum rates benefit the network and provide varied data for subsequent AIGX model training, enabling the model to adapt to diverse network conditions.}
	\label{case}
\end{figure*}

Managing efficient communication between a base station and a user across multiple channels, as depicted in Part B of Fig.~\ref{case}, is a representative challenge in modern wireless systems. 
While conventional techniques like water filling are precise, they are resource-intensive and need constant adjustments for each set of channel gains. On the other hand, AIGX offers an adaptive, real-time solution suitable for fluctuating channel conditions.

The case study\footnote{The code is available on \url{https://github.com/HongyangDu/VirtuousAIGX}.}, with different environmental stages $T_1$, $T_2$, and $T_3$ switching between AIGX model training and channel data collection, shows AIGX's adaptability in optimizing power allocation based on real-time channel feedback. This adaptability suggests AIGX's broader applicability, which could be the key to enhancing a range of wireless communication problems where environmental conditions and data patterns change rapidly, making this case study pivotal in demonstrating the impact of AIGX on network optimization.

\subsection{Initial Conditions and AIGX Training (\(T_1\))}
During \(T_1\), the network operates under stable conditions, analogous to good weather conditions. With \(M = 20\), channel gains are set between \(5\) and \(8\) for the initial \(10\) channels and \(3\) to \(6\) for the subsequent \(10\) to simulate a range of channel conditions. These gains, serving as conditions, are collected in training the GDM model to allocate power by identifying and adapting to channel-specific states to maximize the data rate. Specifically, the GDM model is trained through a process where optimal decision generation is achieved by introducing noise to an expert solution and denoising it, as demonstrated in \textit{Part A}. As shown in Part C of Fig.~\ref{case}, by the close of $T_1$ model training, the AIGX model enhances the network data rate by $18.8\%$ compared with the average allocation scheme.

\subsection{Environmental Variations and Their Impact (\(T_2\))}
In phase $T_2$ channel data collection, the network grapples with varied environmental changes, for instance, a rainstorm, which is particularly challenging for wireless communications due to the scattering and absorption of radio signals at high-frequency bands, e.g., mmWave. Specifically, we consider that channel gains fluctuate capriciously between \(1\) and \(7\) for all \(20\) channels, reflecting real-world scenarios where meteorological shifts evoke unpredictable network behavior. Even though the AIGX model trained under \(T_1\) conditions manifests the robustness, it becomes suboptimal in \(T_2\), evidenced by a conspicuous decline in the data rate. Such a situation motivates a virtuous interactive cycle to gain additional expert solutions to enhance the AIGX model training under diverse conditions.

\subsection{The Virtuous Interactive Cycle in Action (\(T_3\))}
In phase \(T_3\), the synergistic virtuous interactive cycle between AIGX and the network becomes evident. The network, under \(T_2\)'s rain-impacted conditions, actively acquires new channel gains and corresponding expert solutions, serving to retrain the AIGX model. This adaptation of the model’s power allocation strategy to the new channel conditions triggers a significant improvement in the data rate. Specifically, the virtuous gain, defined as the enhancement in data rate achieved by the retrained AIGX model compared to a conventional average method, is \(15.1\%\). Conversely, a deep reinforcement learning-based method. i.e., SAC~\cite{du2023beyond}, cannot achieve performance analogous to AIGX, attaining merely an \(8.9\%\) virtuous gain even under the virtuous interactive cycle.

\subsection{Lesson Learned}
The virtuous gain illuminates the importance of the AIGX-network virtuous interactive cycle, emphasizing the essential role of the network in the AIGX-network cycle. This stresses the importance of ongoing data acquisition and feedback mechanisms in maintaining the relevance and adaptability of AIGX algorithms under different network conditions. Rather than just benefiting from AIGX, networks actively contribute to AIGX adaptability and fortifying decision-making efficacy.

\section{Future Directions}

\subsection{AIGX Enhancing Networks}
AIGX could serve as a foundation for automating complex network management tasks, streamlining data flow, and enhancing security. We discuss some representative future research directions:
\begin{itemize}
    \item \textbf{AIGX-driven Near-field Communications:} The unique challenges in near-field MIMO communications, especially complex antenna dependencies~\cite{liu2023near}, are mitigated by AIGX by leveraging its capability to model and adapt to the multifaceted channel dynamics. AIGX’s learning algorithms adapt to the complex channel states, facilitating optimal signal processing and resource allocation.
    % Near-field MIMO communications encounter distinct challenges contrasting with far-field scenarios, notably the complex dependencies between antennas due to proximity. Integrating AIGX solves these high-dimensional signaling environments by leveraging its capability to model and adapt to the multifaceted channel dynamics. AIGX’s learning algorithms adapt to the complex channel states, facilitating optimal signal processing and resource allocation.
    \item \textbf{AIGX-driven Space-Air-Ground Integrated Network (SAGIN):} AIGX addresses the need for robust communication across different atmospheric layers in SAGIN. By crafting adaptive pathways for data flow among satellite, aerial, and terrestrial layers, AIGX's adaptability meets the network's diverse demands, resulting in stable data transmission.
    \item \textbf{AIGX-driven Multimodal Communications:} For networks handling diverse data types like text, image, and video, AIGX agents ensure synchronization and prioritization. They also facilitate real-time, cross-modal resource management, elevating the user experience.
\end{itemize}

\subsection{Networks Supporting AIGX}
Networks empower AIGX's functionalities by providing robust data transportation, enabling rapid model deployment, and facilitating edge computing capabilities. We identify the following avenues for future research:
\begin{itemize}
    \item \textbf{Swarm Intelligence for Collaborative AIGX Service}: Networks enabled with swarm intelligence can distribute AIGX services across nodes more efficiently. For instance, swarm algorithms could distribute GAI tasks across network nodes and exchange information efficiently, thus accelerating model training and inference.
    \item \textbf{Transfer Learning Across Network Hubs:} Network-supported transfer learning can enhance AIGX efficiency by distributing pre-trained models across multiple hubs. This allows each node to leverage shared learning and insights, reducing the need for AIGX services to train new models from scratch.
    \item \textbf{Energy-Efficient Networking for Sustainable AIGX:} To address the high energy demands of AIGX models like ChatGPT, networks can implement energy-aware routing algorithms and task distribution strategies. Data-intensive AIGX tasks may be routed through network nodes utilizing renewable energy sources, and low-energy hardware accelerators can be engaged for specific computations. Additionally, energy harvesting techniques can be integrated to optimize energy use further.
\end{itemize}

\section{Conclusion}
In this article, we have explored AIGX's role in intelligent networks. Incorporated across different network layers, AIGX promotes adaptive responses and enhances network management. Networks, in return, are instrumental in the AIGX lifecycle, from data collection to reducing inference latency. A case study on AIGX-driven power allocation highlighted the ``virtuous interactive cycle'' between AIGX and networks, emphasizing their reciprocal benefits. While our findings illuminate the promising relationship between AIGX and networks, continued research is essential to develop new methodologies and address emerging challenges, aiming for joint advancement of AIGX and intelligent networks.

\bibliographystyle{IEEEtran}
\bibliography{Ref}
\end{document}